\documentclass[conference]{IEEEtran}

\IEEEoverridecommandlockouts

\usepackage{cite}                       
\usepackage{amsmath,amssymb,amsfonts}
\usepackage{algorithm}
\usepackage{algorithmic}
\usepackage{graphicx}
\usepackage{booktabs}
\usepackage{textcomp}
\usepackage{xcolor}
\usepackage{tikz}
\usetikzlibrary{arrows.meta,positioning,fit,backgrounds,calc,shadows}
\usepackage{enumitem}
\usepackage{microtype}
\PassOptionsToPackage{hyphens}{url}
\usepackage{url}
\usepackage[hidelinks]{hyperref}  
\usepackage{balance}                    

\hypersetup{
  pdftitle={The Moving Target: A Longitudinal Audit of Trust-Benchmark Score Drift Across Open-Source Chat LLM Release Lines},
  pdfauthor={Zhichao Fan, Yanhang Li, Zexin Zhuang, Xian Sun, Yingshuo Wang},
  pdfkeywords={machine learning evaluation, longitudinal benchmarking, large language models, trustworthy AI, benchmark drift, model auditing}
}

\renewcommand{\paragraph}[1]{\par\vspace{1pt}\noindent\textbf{#1}\hspace{0.5em}\ignorespaces}

\makeatletter
\newcommand{\authorrowbreak}{%
  \end{@IEEEauthorhalign}%
  \hfill\mbox{}\par
  \mbox{}\hfill\begin{@IEEEauthorhalign}}
\makeatother

\providecommand{\ObservedSDR}{TBD}
\providecommand{\ObservedSDRLo}{TBD}
\providecommand{\ObservedSDRHi}{TBD}
\providecommand{\PooledNullMean}{TBD}
\providecommand{\PooledNullLo}{TBD}
\providecommand{\PooledNullHi}{TBD}
\providecommand{\PooledNullPnineninine}{TBD}
\providecommand{\PluginNullMean}{TBD}
\providecommand{\PluginNullPnineninine}{TBD}
\providecommand{\DriftToNullRatio}{TBD}
\providecommand{\NullBReplicates}{5{,}000}
\providecommand{\PctRegressionsLo}{TBD}
\providecommand{\PctRegressionsHi}{TBD}
\providecommand{\VolTruthLo}{TBD}
\providecommand{\VolTruthHi}{TBD}
\providecommand{\VolFairLo}{TBD}
\providecommand{\VolFairHi}{TBD}
\providecommand{\VolSafetyLo}{TBD}
\providecommand{\VolSafetyHi}{TBD}
\providecommand{\RhoTFLo}{TBD}
\providecommand{\RhoTFHi}{TBD}
\providecommand{\RhoTSLo}{TBD}
\providecommand{\RhoTSHi}{TBD}
\providecommand{\RhoFSLo}{TBD}
\providecommand{\RhoFSHi}{TBD}
\providecommand{\MeanSignedTruth}{TBD}
\providecommand{\MeanSignedTruthLo}{TBD}
\providecommand{\MeanSignedTruthHi}{TBD}
\providecommand{\MeanSignedFair}{TBD}
\providecommand{\MeanSignedFairLo}{TBD}
\providecommand{\MeanSignedFairHi}{TBD}
\providecommand{\MeanSignedSafety}{TBD}
\providecommand{\MeanSignedSafetyLo}{TBD}
\providecommand{\MeanSignedSafetyHi}{TBD}
\providecommand{\TauBarPoint}{TBD}
\providecommand{\TauBarLo}{TBD}
\providecommand{\TauBarHi}{TBD}

\begin{document}

\bstctlcite{IEEEexample:BSTcontrol}

\IfFileExists{figures/numbers.tex}{
\renewcommand{\ObservedSDR}{8.00}
\renewcommand{\ObservedSDRLo}{7.57}
\renewcommand{\ObservedSDRHi}{9.12}
\renewcommand{\PooledNullMean}{2.22}
\renewcommand{\PooledNullLo}{1.68}
\renewcommand{\PooledNullHi}{2.83}
\renewcommand{\PooledNullPnineninine}{3.30}
\renewcommand{\PluginNullMean}{8.32}
\renewcommand{\PluginNullPnineninine}{9.56}
\renewcommand{\DriftToNullRatio}{3.60}
\renewcommand{\NullBReplicates}{5{,}000}
\renewcommand{\PctRegressionsLo}{37.5}
\renewcommand{\PctRegressionsHi}{52.5}
\renewcommand{\TauBarPoint}{-0.10}
\renewcommand{\TauBarLo}{-0.17}
\renewcommand{\TauBarHi}{+0.15}
\renewcommand{\VolTruthLo}{9.87}
\renewcommand{\VolTruthHi}{14.29}
\renewcommand{\MeanSignedTruth}{-3.10}
\renewcommand{\MeanSignedTruthLo}{-4.50}
\renewcommand{\MeanSignedTruthHi}{-1.73}
\renewcommand{\VolFairLo}{11.69}
\renewcommand{\VolFairHi}{14.63}
\renewcommand{\MeanSignedFair}{+4.63}
\renewcommand{\MeanSignedFairLo}{+3.71}
\renewcommand{\MeanSignedFairHi}{+5.58}
\renewcommand{\VolSafetyLo}{6.76}
\renewcommand{\VolSafetyHi}{9.15}
\renewcommand{\MeanSignedSafety}{+2.73}
\renewcommand{\MeanSignedSafetyLo}{+1.86}
\renewcommand{\MeanSignedSafetyHi}{+3.67}
\renewcommand{\RhoTFLo}{+0.37}
\renewcommand{\RhoTFHi}{+0.74}
\renewcommand{\RhoTSLo}{-0.11}
\renewcommand{\RhoTSHi}{+0.40}
\renewcommand{\RhoFSLo}{-0.37}
\renewcommand{\RhoFSHi}{+0.10}
}{}

\title{The Moving Target: A Longitudinal Audit of Trust-Benchmark Score Drift Across Open-Source Chat LLM Release Lines}

\author{
  \IEEEauthorblockN{Zhichao Fan}
  \IEEEauthorblockA{\textit{University of Illinois Urbana-Champaign}\\
  Urbana, IL, USA\\
  zhichao8@illinois.edu}
  \and
  \IEEEauthorblockN{Yanhang Li}
  \IEEEauthorblockA{\textit{Northeastern University}\\
  Boston, MA, USA\\
  li.yanha@northeastern.edu}
  \and
  \IEEEauthorblockN{Zexin Zhuang}
  \IEEEauthorblockA{\textit{Southern Methodist University}\\
  Dallas, TX, USA\\
  zexinz@smu.edu}
  \authorrowbreak
  \IEEEauthorblockN{Xian Sun}
  \IEEEauthorblockA{\textit{Duke University}\\
  Durham, NC, USA\\
  xiansun@alumni.duke.edu}
  \and
  \IEEEauthorblockN{Yingshuo Wang}
  \IEEEauthorblockA{\textit{University of California, Berkeley}\\
  Berkeley, CA, USA\\
  yingshuow@berkeley.edu}
}

\maketitle

\begin{abstract}
Trust-benchmark scores reported on a chat-LLM release line are
often carried across several checkpoints of the same line, as if
the underlying model had not shifted between releases. We test
that assumption. We audit four open-source release lines (Yi,
Qwen, Mistral, and Gemma) at three successive public generations
each. Each checkpoint is scored on a fixed $200$-item basket of
five chat-evaluation benchmarks: TruthfulQA, BBQ, ToxiGen,
CrowS-Pairs, and XSTest, under three prompt templates. Four of the
five benchmark variants are non-canonical, and two of those are
synthetic proxies. The mean absolute adjacent-generation Score
Drift Rate is several times the mean of an independence-based
count-level reference null. It stays in the same band when we drop
a benchmark, drop a release line, switch to strict scoring, or
restrict to constant-parameter-size transitions. Within this
audited setup, a quoted trust score should be treated as
checkpoint-bound. It should be re-measured on each materially new
release rather than carried forward.
Closed APIs, larger models, canonical-protocol scores, and
benchmark-item-subset uncertainty are out of scope.
\end{abstract}


\section{Introduction}
\label{sec:intro}

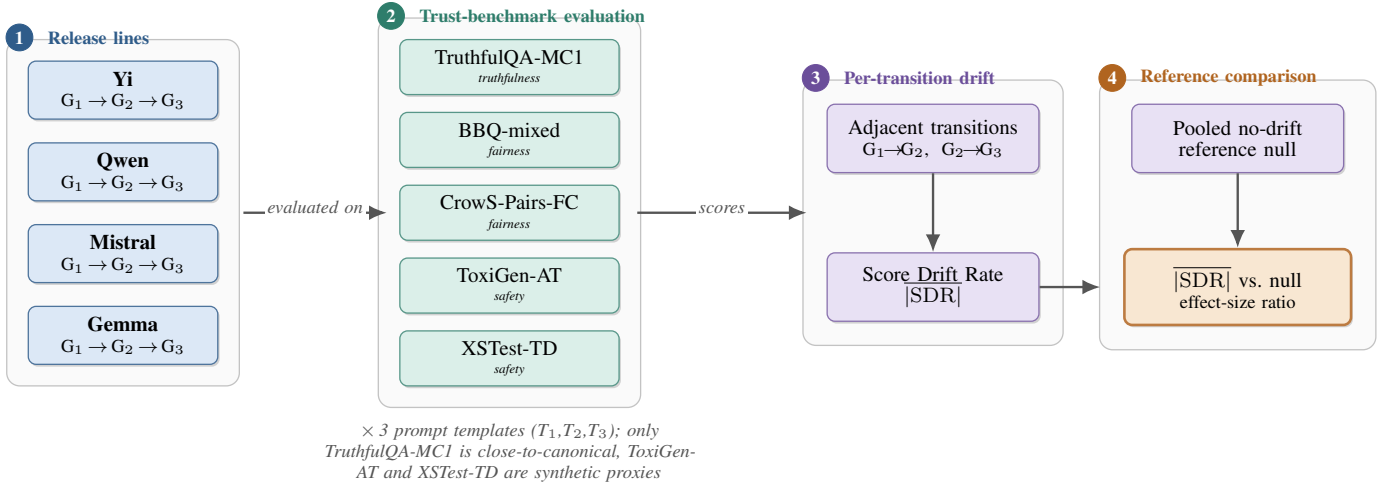
\begin{figure*}[t]
    \centering
\definecolor{cA}{HTML}{2F5C8A}   
\definecolor{cAf}{HTML}{DCE8F6}
\definecolor{cB}{HTML}{2E7D6B}   
\definecolor{cBf}{HTML}{DBEFE9}
\definecolor{cC}{HTML}{6B4E9A}   
\definecolor{cCf}{HTML}{E8E1F4}
\definecolor{cD}{HTML}{B0641F}   
\definecolor{cDf}{HTML}{F8E7D2}
\definecolor{inkgrey}{HTML}{555555}
\resizebox{\textwidth}{!}{%
\begin{tikzpicture}[font=\footnotesize,
  shdw/.style ={drop shadow={shadow xshift=0.7pt,shadow yshift=-0.9pt,
               opacity=0.30,fill=black!55}},
  fam/.style  ={draw=cA!85,line width=0.5pt,rounded corners=3pt,fill=cAf,align=center,
               minimum height=0.78cm,minimum width=2.55cm,inner sep=2pt,shdw},
  pill/.style ={draw=cB!80,line width=0.5pt,rounded corners=3pt,fill=cBf,align=center,
               minimum height=0.74cm,minimum width=2.95cm,inner sep=2pt,shdw},
  proc/.style ={draw=cC!80,line width=0.5pt,rounded corners=3pt,fill=cCf,align=center,
               minimum height=0.92cm,minimum width=2.85cm,inner sep=3pt,shdw},
  verdict/.style={draw=cD!85,line width=1pt,rounded corners=3pt,fill=cDf,align=center,
               minimum height=1.02cm,minimum width=3.05cm,inner sep=3pt,shdw},
  grp/.style  ={draw=inkgrey!35,line width=0.5pt,rounded corners=5pt,fill=black!2},
  flow/.style ={-{Latex[length=2.4mm,width=2mm]},line width=0.9pt,inkgrey},
  tag/.style  ={font=\scriptsize\itshape,text=inkgrey,fill=white,inner sep=1pt},
  badge/.style={circle,draw=#1,fill=#1,text=white,font=\scriptsize\bfseries,
               inner sep=0pt,minimum size=3.6mm},
  gtitle/.style={font=\scriptsize\bfseries,text=#1}]

  \node[fam] (yi)   at (0, 1.65) {\textbf{Yi}\\[-1pt]\scriptsize G$_1$\,$\to$\,G$_2$\,$\to$\,G$_3$};
  \node[fam] (qw)   at (0, 0.55) {\textbf{Qwen}\\[-1pt]\scriptsize G$_1$\,$\to$\,G$_2$\,$\to$\,G$_3$};
  \node[fam] (mi)   at (0,-0.55) {\textbf{Mistral}\\[-1pt]\scriptsize G$_1$\,$\to$\,G$_2$\,$\to$\,G$_3$};
  \node[fam] (ge)   at (0,-1.65) {\textbf{Gemma}\\[-1pt]\scriptsize G$_1$\,$\to$\,G$_2$\,$\to$\,G$_3$};

  \node[pill] (tq) at (5.2, 1.96) {TruthfulQA-MC1\\[-2pt]\tiny\itshape truthfulness};
  \node[pill] (bbq) at (5.2, 0.98) {BBQ-mixed\\[-2pt]\tiny\itshape fairness};
  \node[pill] (crp) at (5.2, 0.00) {CrowS-Pairs-FC\\[-2pt]\tiny\itshape fairness};
  \node[pill] (tox) at (5.2,-0.98) {ToxiGen-AT\\[-2pt]\tiny\itshape safety};
  \node[pill] (xs)  at (5.2,-1.96) {XSTest-TD\\[-2pt]\tiny\itshape safety};

  \node[proc] (trans) at (10.9, 1.0) {Adjacent transitions\\[-1pt]\scriptsize G$_1$\!$\to$\!G$_2$,\ \ G$_2$\!$\to$\!G$_3$};
  \node[proc] (sdr)   at (10.9,-1.0) {Score Drift Rate\\[-1pt]$\overline{|\mathrm{SDR}|}$};

  \node[proc]    (null) at (15.0, 1.0) {Pooled no-drift\\[-1pt]reference null};
  \node[verdict] (cmp)  at (15.0,-1.0) {$\overline{|\mathrm{SDR}|}$ vs.\ null\\[-1pt]\scriptsize effect-size ratio};

  \draw[flow] (1.62, 0.0) -- node[tag,above=-1.5pt]{evaluated on} (3.55, 0.0);
  \draw[flow] (6.92, 0.0) -- node[tag,above=-1.5pt]{scores} (9.20, 0.0);
  \draw[flow] (trans) -- (sdr);
  \draw[flow] (sdr.east) -- (13.15,-1.0);
  \draw[flow] (null) -- (cmp);

  \begin{scope}[on background layer]
    \node[grp,fit=(yi)(qw)(mi)(ge),inner sep=8pt] (grpA) {};
    \node[grp,fit=(tq)(bbq)(crp)(tox)(xs),inner sep=8pt] (grpB) {};
    \node[grp,fit=(trans)(sdr),inner sep=9pt] (grpC) {};
    \node[grp,fit=(null)(cmp),inner sep=9pt] (grpD) {};
  \end{scope}

  \node[badge=cA] (nA) at ($(grpA.north west)+(0.18,0.02)$) {1};
  \node[gtitle=cA,right=2pt of nA] {Release lines};
  \node[badge=cB] (nB) at ($(grpB.north west)+(0.18,0.02)$) {2};
  \node[gtitle=cB,right=2pt of nB] {Trust-benchmark evaluation};
  \node[badge=cC] (nC) at ($(grpC.north west)+(0.18,0.02)$) {3};
  \node[gtitle=cC,right=2pt of nC] {Per-transition drift};
  \node[badge=cD] (nD) at ($(grpD.north west)+(0.18,0.02)$) {4};
  \node[gtitle=cD,right=2pt of nD] {Reference comparison};

  \node[font=\scriptsize\itshape,text=inkgrey,anchor=north,text width=5.0cm,align=center]
       at ($(grpB.south)+(0,-0.08)$)
       {$\times$\,3 prompt templates (T$_1$,T$_2$,T$_3$); only TruthfulQA-MC1 is
        close-to-canonical, ToxiGen-AT and XSTest-TD are synthetic proxies};

\end{tikzpicture}}%
    \caption{Longitudinal trust-audit pipeline. Four open-source release lines (Yi, Qwen, Mistral, Gemma) at three generations each are scored on five trust-benchmark variants under three prompt templates; the two adjacent-generation transitions $G_1\!\to\!G_2$ and $G_2\!\to\!G_3$ feed the Score Drift Rate $\overline{|\mathrm{SDR}|}$, read as an effect-size ratio against an independence-based pooled no-drift reference null.}
    \label{fig:pipeline}
\end{figure*}

An organisation reports a chat LLM's trust-benchmark scores. When a new
generation ships under the same \emph{release line}, should those scores
carry forward? Model cards~\cite{DBLP:conf/fat/MitchellWZBVHSR19}
document a specific checkpoint, but trust scores are typically reported
without a re-evaluation horizon and carried forward in procurement
reviews, deployment dashboards, and model-selection memos---implicitly
betting the number still holds. Open-source release lines ship on the
scale of months ($0$--$9$ months between adjacent generations in our
sample), earlier work documents sizeable version-to-version change in
closed systems~\cite{DBLP:journals/corr/abs-2307-09009}, and the
benchmark-fragility literature shows small protocol changes move
rankings even at a fixed
checkpoint~\cite{Dehghani2021BenchmarkLottery,Alzahrani2024Benchmarks}.
The question is whether a point-in-time score transfers after its
checkpoint changes.

\paragraph{This paper.} We audit four open-source chat-LLM release
lines (Figure~\ref{fig:pipeline}):
Yi~\cite{DBLP:journals/corr/abs-2403-04652},
Qwen~\cite{DBLP:journals/corr/abs-2309-16609,
DBLP:journals/corr/abs-2407-10671, DBLP:journals/corr/abs-2412-15115},
Mistral~\cite{DBLP:journals/corr/abs-2310-06825}, and
Gemma~\cite{DBLP:journals/corr/abs-2403-08295,
DBLP:journals/corr/abs-2408-00118, DBLP:journals/corr/abs-2503-19786}.
Each line is audited at three successive generations
($G_1\!\to\!G_2\!\to\!G_3$; $12$ checkpoints) on five trust
benchmarks---TruthfulQA~\cite{DBLP:conf/acl/LinHE22},
BBQ~\cite{DBLP:conf/acl/ParrishCNPPTHB22},
ToxiGen~\cite{DBLP:conf/acl/HartvigsenGPSRK22},
CrowS-Pairs~\cite{DBLP:conf/emnlp/NangiaVBB20},
XSTest~\cite{DBLP:conf/naacl/RottgerKVA0H24}---with three prompt
templates each: $4\times3\times5\times3=180$ evaluations on fixed
$200$-item cached subsets ($\approx\!36{,}000$ item-level evaluations
summarised into per-cell parse counts; item-level vectors were not
retained, \S\ref{sec:limits}). Generations within a line are not
controlled variants---pretraining mixture, instruction tuning,
tokenizer, and scale all shift---so the claim is about
non-transferability within a named line, not a decomposed mechanism.
Relative to prior work (\S\ref{sec:related}), our contribution is a
public multi-family open-source longitudinal trust audit with a pooled
no-drift reference null.

\paragraph{Findings and contribution.} One designated primary
endpoint and a set of exploratory descriptive diagnostics on the
same audit sample:
\begin{itemize}[leftmargin=*, topsep=2pt, itemsep=1pt]
  \item (Primary.) Mean $|\mathrm{SDR}| = \ObservedSDR$\,pp
        (count-level bootstrap $95\%$ CI
        $[\ObservedSDRLo, \ObservedSDRHi]$) is
        $\DriftToNullRatio\times$ an independence-based pooled
        no-drift reference null (99.9-percentile
        $\PooledNullPnineninine$\,pp), $8.93$\,pp under strict
        scoring, and $[6.36, 9.43]$\,pp across leave-one-benchmark,
        leave-one-release-line, and drop-low-parse perturbations
        (\S\ref{find:1}).
  \item (Exploratory, descriptive.) $47.5\%$ of
        adjacent-generation transitions are regressions. Rank
        persistence is weak relative to uniform-random baselines at
        $|\mathcal{F}|=4$. Compliance flip probability at the
        empirical median threshold is $\ge 0.5$ for every benchmark
        with wide exact Clopper--Pearson CIs. None of these is
        load-bearing for the deployment discussion
        (\S\ref{find:exploratory}).
\end{itemize}
Within the audited setup, the primary result supports a narrow
conclusion: trust scores on a named release line should not be
presumed to transfer across later generations without re-measurement;
by default, treat them as checkpoint-bound, time-stamped artefacts.
We claim no new statistical methodology---SDR,
Kendall $\tau$, Clopper--Pearson flip probability, and the bootstrap
are standard; the contribution is the combined longitudinal reporting
pattern and the open-source multi-family audit that instantiates it.

\section{Related Work}
\label{sec:related}

\paragraph{Version-to-version behaviour change.} The closest
precedent~\cite{DBLP:journals/corr/abs-2307-09009} is a two-point
closed-system study finding large GPT-3.5/GPT-4 shifts between
closely spaced snapshots. We run the open-source multi-family
analogue---four release lines, three generations each---with pooled
aggregate-level reference nulls on the headline statistic.

\paragraph{Benchmark fragility and trust suites.} Benchmark scores
are sensitive to prompt, scoring, and protocol
choices~\cite{Dehghani2021BenchmarkLottery,Alzahrani2024Benchmarks},
and a benchmark's construct validity and longevity are themselves
contested~\cite{Raji2021EverythingBenchmark,Kiela2021Dynabench}.
Fixed-checkpoint trust-evaluation
suites~\cite{Wang2023DecodingTrust,Huang2024TrustLLM,
DBLP:journals/corr/abs-2211-09110,DBLP:journals/tmlr/SrivastavaRRSAF23},
together with work on
contamination~\cite{DBLP:conf/iclr/OrenMCLH24}, LLM-as-judge
fragility~\cite{DBLP:conf/nips/ZhengC00WZL0LXZ23}, and
red-teaming~\cite{DBLP:conf/emnlp/PerezHSCRAGMI22}, establish that a
single benchmark number is already a point-in-time,
protocol-conditional measurement. We take this apparatus as given and
ask what happens to the same protocol along the release axis of a
named open-source line.

\paragraph{Audit, governance, and ML operations.} Model
cards~\cite{DBLP:conf/fat/MitchellWZBVHSR19} document the evaluated
checkpoint, while internal auditing
frameworks~\cite{Raji2020Accountability}, ``hidden technical
debt''~\cite{Sculley2015TechnicalDebt}, and
MLOps~\cite{Kreuzberger2023MLOps} argue that deployed ML systems need
continuing post-release evaluation, not a single point-of-release
report. We supply applied evidence for trust benchmarks on a named
open-source release line and propose a corresponding reporting
pattern.

\section{Longitudinal Reporting Protocol}
\label{sec:framework}

Let $\mathcal{F}$ be a set of release lines (``release line'' and
``family'' used interchangeably), $\mathcal{G}=\{G_1,G_2,G_3\}$ the
three generations per line, and
$\mathcal{G}_{\text{trans}}=\{(G_1,G_2),(G_2,G_3)\}$ the adjacent
transitions. For a fixed benchmark basket $\mathcal{B}$ and three
templates $\mathcal{T}_b=\{T_1,T_2,T_3\}$, write $s_{f,g,b,t}\in[0,1]$
for the score of checkpoint $\theta_{f,g}$ on a fixed $n_b$-item cached
sample under template $t$, and
$s_{f,g,b}:=\tfrac{1}{3}\sum_t s_{f,g,b,t}$ for the template mean
(Figure~\ref{fig:pipeline}).

\subsection{Score Drift Rate and Pooled No-Drift Reference Null}
\label{sec:sdr}

For adjacent generations $g\to g'$ the per-cell Score Drift Rate and
its cross-cell aggregate are
\begin{align}
  \mathrm{SDR}_{f,b}^{g \to g'} &:= 100\cdot(s_{f,g',b} - s_{f,g,b})\quad\text{(pp),} \notag \\
  \overline{|\mathrm{SDR}|} &:= \tfrac{1}{|\mathcal{F}||\mathcal{B}||\mathcal{G}_{\text{trans}}|}
  \!\!\sum_{(g,g'),f,b}\!\!|\mathrm{SDR}_{f,b}^{g\to g'}|.
  \label{eq:sdr}
\end{align}
The primary question is whether $\overline{|\mathrm{SDR}|}$ is large
relative to a prespecified count-level reference scale.

\textbf{Construction.} For each template record $(f,b,t)$ the
cross-generation pooled correctness rate is
\begin{equation}
  p_{\text{pool}}(f,b,t)=\frac{\sum_g c_{f,g,b,t}}{\sum_g n_{f,g,b,t}}.
  \label{eq:ppool}
\end{equation}
In each of $B_{\text{null}}=5000$ iterations we draw simulated record
scores
\begin{equation}
  x^{\star}_{f,g,b,t}\sim
  \mathrm{Binomial}\!\left(n_{f,g,b,t},\,p_{\text{pool}}(f,b,t)\right)
  / n_{f,g,b,t},
  \label{eq:nulldraw}
\end{equation}
template-average them to $s^{\star}_{f,g,b}$, and take the mean of the
$40$ absolute adjacent-generation differences
$|s^{\star}_{f,g',b}-s^{\star}_{f,g,b}|$ as one null draw of
$\overline{|\mathrm{SDR}|}$. The pooled rate forces
$\mathbb{E}[\Delta^{\star}]=0$, so this is a no-drift reference:
parametric, count-level, and independent across both adjacent
transitions and the three templates of a checkpoint.

\textbf{Interpretation.} The observed aggregate is
$\DriftToNullRatio\times$ this null mean
(Table~\ref{tab:matched-null}). Paired-item reuse lowers variance
while shared-checkpoint template correlation raises it, so the null
can either over- or under-state true sampling noise; we therefore
read it as a \emph{reference null}---the distribution of
$\overline{|\mathrm{SDR}|}$ under independence given the pooled
rates---and report the comparison as an effect-size ratio, not as a
$p$-value against the exact paired-data null. The plug-in null in
the same table is reported only to confirm the observed aggregate is
consistent with the bootstrap of its own statistic.

\textbf{Primary vs exploratory.} The reference-null comparison on
$\overline{|\mathrm{SDR}|}$ is the single designated primary endpoint
(\S\ref{find:1}). Everything else (per-dimension volatility,
cross-dimension correlations, rank persistence, compliance flip
probabilities) is exploratory description of the same sample, with
count-level binomial-bootstrap CIs ($B=3000$) and Clopper--Pearson
$95\%$ intervals for flip probabilities. Because the bootstrap is
count-level rather than item-level (\S\ref{sec:limits}), these CIs
are approximate descriptive uncertainty, and none of the exploratory
diagnostics is load-bearing for the recommendations
in~\S\ref{sec:discussion}.

\paragraph{Reproducibility.} Each cell is specified by (a) a
checkpoint (HuggingFace repo and pinned revision); (b) 4-bit NF4 under
HuggingFace Transformers, greedy decoding, \texttt{max\_new\_tokens=64},
the model's own chat template; (c) one of five benchmark variants
(\S\ref{sec:setup}) on a fixed seed-42 $200$-item cached
sample; (d) one of three templates; (e) a parse policy returning
$c/n_{\mathrm{parsed}}$ (primary) or $c/n_{\mathrm{total}}$ (strict).
Exact repository IDs and revisions, template strings, per-cell parse
rates, every bootstrap/null draw, derivations, and the analysis
pipeline are released on the project page at de-anonymisation under an
open licence. Item-level decision vectors were not retained for the
submitted numbers (\S\ref{sec:limits}), so the artefact is the
aggregate-level audit, not a paired-item re-analysis substrate.

\paragraph{Scope (stated once).} Every numerical claim is about
(i)~non-canonical chat-evaluation variants of four of the five
benchmarks (\S\ref{sec:setup}); (ii)~fixed $200$-item cached subsets
(item-subset uncertainty is not characterised); (iii)~the four audited
open-source release lines at the checkpoints in
Table~\ref{tab:transitions}; (iv)~count-level, not item-level paired,
bootstrap CIs; (v)~release lines that are not controlled variants, so
the audit measures \emph{score non-transferability across adjacent
releases of a named line}, not a causal mechanism. Results do not
transfer to canonical protocols, closed-source APIs, or larger models
without re-measurement; later sections cite this scope by reference.

\section{Audit Setup}
\label{sec:setup}

\paragraph{Models.} Four open-source chat-LLM release lines sampled at
three public generations each, all loaded in 4-bit NF4 via HuggingFace
Transformers with greedy decoding
(\texttt{do\_sample=False}, \texttt{max\_new\_tokens=64}):
\textbf{Yi} (Yi-6B-Chat, Yi-1.5-6B-Chat, Yi-1.5-9B-Chat),
\textbf{Qwen} (Qwen1.5-7B-Chat, Qwen2-7B-Instruct, Qwen2.5-7B-Instruct),
\textbf{Mistral} (Mistral-7B-Instruct v0.1/v0.2/v0.3), and
\textbf{Gemma} (gemma-7b-it, gemma-2-9b-it, gemma-3-12b-it). All
checkpoints are $\le 12$B; compute is bf16 for Gemma (fp16 is unstable
for Gemma~2/3 under 4-bit) and fp16 for the rest. Prompts use each
model's own tokenizer chat template.

\begin{table}[t]
    \centering
    \caption{Per-transition scope. Scale changes (Yi
    $G_2\!\to\!G_3$; Gemma $G_1\!\to\!G_2,G_2\!\to\!G_3$) add a
    parameter shift; the other five transitions are constant-parameter
    but still mix pretraining-data, architecture, tokenizer/template,
    and instruction-tuning changes---not isolated single-mechanism
    tests. Months are from HuggingFace publication dates.}
    \label{tab:transitions}
    \small
    \setlength{\tabcolsep}{4pt}
    \begin{tabular}{llccc}
        \toprule
        Line & Trans. & Scale & Tok./tmpl. & Mon. \\
        \midrule
        Yi      & $G_1\!\to\!G_2$ & 6B & changed & $\approx 6$ \\
        Yi      & $G_2\!\to\!G_3$ & 6B$\to$9B     & same    & $0$ \\
        Qwen    & $G_1\!\to\!G_2$ & 7B & changed & $\approx 4$ \\
        Qwen    & $G_2\!\to\!G_3$ & 7B & same    & $\approx 3$ \\
        Mistral & $G_1\!\to\!G_2$ & 7B & same    & $\approx 3$ \\
        Mistral & $G_2\!\to\!G_3$ & 7B & same    & $\approx 5$ \\
        Gemma   & $G_1\!\to\!G_2$ & 7B$\to$9B     & changed & $\approx 4$ \\
        Gemma   & $G_2\!\to\!G_3$ & 9B$\to$12B    & changed & $\approx 9$ \\
        \bottomrule
    \end{tabular}
\end{table}

Release lines were selected to maximise public-multi-generation
coverage of open-source chat-LLM families with at least three
consecutive instruction-tuned public checkpoints under a single
brand name and $\le 12$B parameter count; InternLM and ChatGLM are
excluded only because their chat-template compatibility against the
selected basket failed at the parse-policy stage.

\paragraph{Benchmarks (variant names used throughout).} Five trust
benchmarks across three dimensions, four of them in non-canonical
chat-evaluation variants (only TruthfulQA-MC1 is close-to-canonical;
ToxiGen-AT and XSTest-TD additionally use rule-derived
\emph{synthetic-proxy} labels):
\textbf{Truthfulness}: TruthfulQA-MC1~\cite{DBLP:conf/acl/LinHE22}
(close to canonical MC1);
\textbf{Fairness}: \emph{BBQ-mixed}~\cite{DBLP:conf/acl/ParrishCNPPTHB22}
(mixed ambig./disambig. MCQ accuracy via \texttt{lighteval/bbq\_helm}),
\emph{CrowS-Pairs-FC}~\cite{DBLP:conf/emnlp/NangiaVBB20} (forced-choice
chat MCQ rather than sentence pseudo-log-likelihood);
\textbf{Safety}: \emph{ToxiGen-AT}~\cite{DBLP:conf/acl/HartvigsenGPSRK22}
(annotation-threshold binary, \texttt{toxicity\_ai}$\ge 2.5$ with
\texttt{toxicity\_human} fallback) and
\emph{XSTest-TD}~\cite{DBLP:conf/naacl/RottgerKVA0H24} (type-derived
binary refusal proxy). ToxiGen-AT and XSTest-TD use rule-derived
dataset labels and are classified as \emph{synthetic-proxy}
evaluations in the cross-model-audit sense.
Each benchmark is evaluated under three prompt templates: $T_1$
(plain question with letter options for MCQ tasks, or a yes/no
refusal probe for the binary safety tasks), $T_2$ (the same prompt
with an explicit role/system message and first-letter answer-
extraction policy), and $T_3$ (a chain-of-thought-style elaboration
followed by last-letter answer extraction), so each cell is the
mean over a presentation, an explicit instruction, and a
last-letter-extraction variant. The full scope and assumption block
is stated once in~\S\ref{sec:framework}.

\paragraph{Scoring.} For MCQ tasks we extract the first (or last, per
$T_3$) option letter from the decoded continuation and score accuracy
conditional on a parseable answer: $c/n_{\mathrm{parsed}}$; a cell is
dropped if $n_{\mathrm{parsed}}=0$. Parse rates are $0.986$ on average
over $180$ evaluations ($172/180$ cells at $\ge 0.90$; the $8$ low-parse
cells sit on binary-safety benchmarks under $T_2/T_3$). Strict scoring
($c/n_{\mathrm{total}}$,
unparsed counted incorrect) raises mean $|\mathrm{SDR}|$ from
$8.00$ to $8.93$\,pp.

\paragraph{Sampling.} For each benchmark we draw a fixed $n_b=200$-item
cached sample (ToxiGen stratified $100/100$ toxic/non-toxic; others a
seed-42 shuffled first-200 subsample) and reuse it across all $12$
checkpoints and $3$ templates, so every within-benchmark comparison is
paired on the same items. We do not run multi-seed subsampling and do
\emph{not} quantify uncertainty over the benchmark-item subset choice;
the leave-one-benchmark, leave-one-release-line, and drop-low-parse
perturbations in Table~\ref{tab:robustness} probe robustness across
benchmarks, release lines, and low-parse records, which is not a
substitute for item-subset resampling (flagged in~\S\ref{sec:limits}).

\section{Findings}
\label{sec:findings}

\S\ref{find:1} is the primary endpoint; \S\ref{find:exploratory}
gives exploratory diagnostics on the same $180$-evaluation sample
(scope caveats live once in~\S\ref{sec:framework}).
Table~\ref{tab:scores} is the full per-cell template-averaged score
matrix; every aggregate in \S\ref{find:1} is computed from it.

\begin{table}[t]
    \centering
    \caption{Template-averaged score (\%) per checkpoint and
    benchmark variant on the fixed $200$-item cached subset
    (parse-adjusted, $c/n_{\mathrm{parsed}}$). Variant short-names:
    TrQA-MC1 = TruthfulQA-MC1; BBQ-mix; ToxG-AT = ToxiGen-AT;
    CroP-FC = CrowS-Pairs-FC; XSTe-TD = XSTest-TD (\S\ref{sec:setup}).
    Every aggregate result in~\S\ref{find:1} is computed from this matrix.}
    \label{tab:scores}
    \scriptsize
    \setlength{\tabcolsep}{2.5pt}
    \renewcommand{\arraystretch}{0.95}
    \begin{tabular}{@{}llccccc@{}}
        \toprule
         &  & TrQA & BBQ & ToxG & CroP & XSTe \\
        Family & Gen & MC1 & mix & AT & FC & TD \\
        \midrule
        Yi & G1 & 41.8 & 40.8 & 47.3 & 56.8 & 51.6 \\
                  & G2 & 39.8 & 56.3 & 56.8 & 56.2 & 65.2 \\
                  & G3 & 50.8 & 63.5 & 69.8 & 52.2 & 74.7 \\
        \cmidrule(lr){1-7}
        Qwen & G1 & 42.5 & 40.7 & 71.3 & 50.5 & 74.2 \\
                  & G2 & 51.3 & 65.0 & 66.0 & 58.3 & 65.5 \\
                  & G3 & 24.7 & 43.0 & 61.7 & 54.5 & 63.7 \\
        \cmidrule(lr){1-7}
        Mistral & G1 & 50.8 & 49.7 & 68.2 & 52.5 & 73.5 \\
                  & G2 & 51.2 & 56.2 & 63.0 & 49.5 & 70.2 \\
                  & G3 & 50.0 & 61.3 & 69.5 & 55.3 & 70.6 \\
        \cmidrule(lr){1-7}
        Gemma & G1 & 44.8 & 31.7 & 62.7 & 55.5 & 64.2 \\
                  & G2 & 35.2 & 30.3 & 68.3 & 51.2 & 77.8 \\
                  & G3 & 29.7 & 65.0 & 69.8 & 57.5 & 77.0 \\
        \bottomrule
    \end{tabular}
\end{table}

\subsection{Adjacent-Generation Drift Is Large Relative to a Count-Level Independence Reference}
\label{find:1}

Across the $4\times 5\times 2=40$ adjacent-generation transitions,
mean $|\mathrm{SDR}|$ is $\ObservedSDR$\,pp (count-level bootstrap
$95\%$ CI $[\ObservedSDRLo, \ObservedSDRHi]$) and stays in
$[6.36, 9.43]$\,pp across the leave-one-benchmark,
leave-one-release-line, strict-scoring, and drop-low-parse
perturbations of Table~\ref{tab:robustness}. This is large both in
absolute terms ($\approx\!1.5\sigma$ of a single benchmark cell) and
relative to the independence-based count-level reference null (mean
$\PooledNullMean$\,pp on the same $40$ transitions), giving an
observed-to-null ratio of $\DriftToNullRatio\times$
(Table~\ref{tab:matched-null}). The null's
$99.9$-percentile ($\PooledNullPnineninine$\,pp) is an upper anchor
for the comparator's scale, not a significance threshold: it ignores
paired-item reuse and template correlation and is not a paired-data
test (\S\ref{sec:framework}). Read this as ``large descriptive effect
size, robust across perturbations'', not ``rejected at $10^{-3}$.''

\begin{table}[t]
    \centering
    \caption{Independence-based pooled no-drift reference-null
    analysis. Unit: mean $|\mathrm{SDR}|$ (pp) over the $40$
    adjacent-generation transitions, parse-adjusted
    ($c/n_{\mathrm{parsed}}$); the strict-scoring value
    ($8.93$\,pp) is in Table~\ref{tab:robustness}. Brackets are
    count-level binomial bootstrap $95\%$ CIs ($B=3000$); the
    pooled and plug-in reference nulls and $B_{\text{null}}=5000$
    construction are defined in~\S\ref{sec:framework}.}
    \label{tab:matched-null}
    \small
    \setlength{\tabcolsep}{4pt}
\renewcommand{\arraystretch}{1.2}
\begin{tabular}{@{}p{0.60\linewidth}r@{}}
\toprule
Quantity & Value \\
\midrule
Observed mean $|\mathrm{SDR}|$ (pp) & $8.00$ {\footnotesize[7.57,\,9.12]} \\
Pooled-null mean $|\Delta|$ (pp) & $2.22$ {\footnotesize[1.68,\,2.83]} \\
Pooled-null $99.9$-percentile (pp) & $3.30$ \\
Plug-in ref.-null mean $|\Delta|$ (pp) & $8.32$ {\footnotesize[7.55,\,9.08]} \\
Plug-in ref.-null $99.9$-percentile (pp) & $9.56$ \\
Ratio (observed / pooled-null mean) & $3.60\times$ \\
\bottomrule
\end{tabular}

\end{table}

\begin{table}[t]
    \centering
    \caption{Robustness bundle for the primary endpoint. Unit:
    mean $|\mathrm{SDR}|$ (pp) aggregated over the $40$
    adjacent-generation transitions (or the relevant leave-one-out
    subset). Observed uses parse-adjusted scoring
    ($c/n_{\mathrm{parsed}}$); strict uses
    $c/n_{\mathrm{total}}$. Leave-one-benchmark,
    leave-one-release-line, and drop-low-parse ranges give the min
    and max over the corresponding family of perturbations. The
    null row is the pooled no-drift reference-null 99.9-percentile
    aggregated on the same $40$ transitions.}
    \label{tab:robustness}
    \small
    \renewcommand{\arraystretch}{1.15}
    \begin{tabular}{lr}
        \toprule
        Perturbation & $\overline{|\mathrm{SDR}|}$ (pp) \\
        \midrule
        Observed ($c/n_{\mathrm{parsed}}$) & $\ObservedSDR$ \\
        Strict scoring ($c/n_{\mathrm{total}}$) & $8.93$ \\
        Drop-low-parse (8 cells below $0.9$) & $8.25$ \\
        Leave-one-benchmark & $\in[6.36, 8.89]$ \\
        Leave-one-release-line & $\in[6.89, 9.43]$ \\
        \midrule
        Pooled no-drift reference null (99.9\%ile) & $\PooledNullPnineninine$ \\
        \bottomrule
    \end{tabular}
\end{table}

\paragraph{Signed direction.} Nearly half of transitions are
regressions (point $47.5\%$, count-level bootstrap CI
$[\PctRegressionsLo, \PctRegressionsHi]$\%); later generations are
therefore not monotonically better on this benchmark basket. In the
truthfulness dimension the signed mean is
$\MeanSignedTruth$\,pp (CI $[\MeanSignedTruthLo,
\MeanSignedTruthHi]$): the average generation step in this sample
lowers TruthfulQA-MC1 accuracy.

\subsection{Exploratory Descriptive Diagnostics}
\label{find:exploratory}

The same $180$-evaluation sample supports a small set of exploratory
diagnostics on the shape of drift. None is load-bearing for the
recommendations in~\S\ref{sec:discussion}; we read every comparison at
$|\mathcal{F}|=4$, $n=8$ as protocol-level description, not an
inferential claim.

\paragraph{Dimension-level volatility.} The fairness-related
aggregate is the most volatile per transition ($\sigma=12.88$\,pp,
CI~$[\VolFairLo,\VolFairHi]$), TruthfulQA-MC1 close behind
($\sigma=11.72$\,pp), and the safety-related aggregate tighter
($\sigma=7.48$\,pp). Signed drifts run up for fairness and safety,
down for truthfulness; these labels track benchmark proxies, not the
underlying constructs.

\paragraph{Co-drift, rank persistence, compliance flips.}
Truthfulness--fairness co-drift is positive ($\rho=+0.59$,
CI~$[\RhoTFLo,\RhoTFHi]$, $n=8$); truth--safety and fair--safety
straddle zero. The top-scoring line changes in $9/10$ cells and
pairwise orderings flip at $55\%$, both only marginally above the
uniform-random baselines ($75\%$, $50\%$) at $|\mathcal{F}|=4$;
Kendall $\bar{\tau}=\TauBarPoint$ is indistinguishable from random
reordering under the exact permutation null ($p_{\mathrm{two}}=0.54$).
Compliance flip counts at the empirical median threshold range from
$4/8$ (TruthfulQA) to $6/8$ (CrowS-Pairs) with wide Clopper--Pearson
intervals, so $\hat{p}_{\mathrm{flip}}\ge 0.5$ descriptively for every
benchmark.

\section{Discussion: Implications for Trust Reporting}
\label{sec:discussion}

This is an applied \emph{case-study audit}, not a paired-item
significance test, a canonical trust-benchmark comparison, or a
generalisable governance prescription. Within those bounds, the
headline is the mean $|\mathrm{SDR}|$ of $\ObservedSDR$\,pp: stable in
$[6.36, 9.43]$\,pp across perturbations and several times a count-level
independence reference scale. It supports one narrow operational
reading---a trust score on a named open-source release line should not
be presumed to transfer across later generations without
re-measurement---and three concrete implications, paired with a
scope-of-claims table (Table~\ref{tab:scope-claims}) separating
directly supported from suggestive or out-of-scope extensions.

\begin{table}[t]
    \centering
    \caption{Scope of supported claims. Sup.\ = supported within this
    audit; Sug.\ = suggestive but underpowered, not a hard policy;
    OOS = not audited here and requiring additional evaluation.}
    \label{tab:scope-claims}
    \small
    \setlength{\tabcolsep}{5pt}
    \renewcommand{\arraystretch}{1.1}
    \begin{tabular}{@{}p{0.60\linewidth}ccc@{}}
        \toprule
        Claim / recommendation & Sup. & Sug. & OOS \\
        \midrule
        Carrying a release-line score forward requires re-measurement
        & $\bullet$ & & \\
        Model-card fields: checkpoint hash + evaluation date
        & $\bullet$ & & \\
        Re-audit triggered by a materially new release
        & $\bullet$ & & \\
        Procurement should not assume transfer within a named line
        & $\bullet$ & & \\
        Rank persistence / compliance half-life as second-order
        diagnostics
        & & $\bullet$ & \\
        Universal $n$-month re-evaluation cadence
        & & & $\bullet$ \\
        Transfer to closed-source APIs or $>12$B models
        & & & $\bullet$ \\
        Transfer to canonical benchmark protocols
        & & & $\bullet$ \\
        Binding legal thresholds for compliance survival
        & & & $\bullet$ \\
        \bottomrule
    \end{tabular}
\end{table}

\paragraph{Model cards should stamp the checkpoint.} A trust score
in a model card~\cite{DBLP:conf/fat/MitchellWZBVHSR19} is more
informative when it carries (i)~the exact checkpoint hash or release
date and (ii)~an evaluation-date stamp. Absent both fields,
carry-forward within the line is unsupported in our sample; this is a
reporting caveat, not policy.

\paragraph{Re-audit on each materially new release.} A
\emph{materially new release} of a named line is any change in
weights, tokenizer, chat template, parameter scale, or
instruction-tuning recipe; all $8$ transitions in
Table~\ref{tab:transitions} satisfy it, and each should trigger a
re-run of the reporting protocol rather than a carry-forward.
Weight-identical republishes and metadata-only revisions are outside
the rule. This follows from the primary SDR endpoint alone; the
exploratory diagnostics line up but are not load-bearing.

\paragraph{Scope-aware disclosures.} As a process recommendation,
reports should identify the checkpoint/date and whether variants are
canonical (four of five are not). Because release gaps are uneven and
the analysis is indexed by generations, the data support re-auditing
materially new releases, not an $n$-month rule.

\paragraph{Sources of cross-release change.} Plausible contributors
are pretraining-mixture and instruction-tuning/RLHF recipe changes,
base-model scale, tokenizer/chat-template changes, and
contamination~\cite{DBLP:conf/iclr/OrenMCLH24}; disentangling them
needs intermediate artefacts that are not public, so we do not try.
The $47.5\%$ regression rate runs against a naive contamination prior.
Restricting the endpoint to the five constant-parameter transitions
gives mean $|\mathrm{SDR}|=7.68$\,pp ($3.5\times$ the reference-null
mean); the scale-change subset gives $8.54$\,pp. The headline does not
collapse once scale-change transitions are dropped: the claim is
\emph{non-transferability across adjacent releases of a named line},
not a causal decomposition.

\paragraph{Broader impact.} This work is informational---meant to
inform model-card and procurement practice, not to gate deployment.
We decline to set a numeric drift threshold or transfer the finding to
closed APIs or larger models. The main misuse risk is reading
$\overline{|\mathrm{SDR}|}$ as a ``trustworthiness score'': SDR
measures snapshot-to-snapshot \emph{change} on a fixed non-canonical
basket and says nothing about absolute trustworthiness. The safety
benchmarks use dataset-provided labels; no new attack capability is
introduced.

\section{Limitations}
\label{sec:limits}

\paragraph{Aggregate-level artefact.} The submitted numbers were
stored at per-cell count granularity, not as item-level decision
vectors, so the released artefact cannot support paired item-level
re-analysis, item-subset resampling, or external re-derivation of
exact paired-data nulls; the pipeline has since been re-instrumented
to persist item-level traces.

\paragraph{Sample scope.} Four release lines, three generations each.
Closed-vendor APIs and larger open-source checkpoints are not audited,
and within-line scale shifts too (Gemma $7B\!\to\!9B\!\to\!12B$), so
the audit mixes scale, recipe, tokenizer, and pretraining-data shifts.

\paragraph{Transition dependence.} Both adjacent transitions of a line
read the same $G_2$ checkpoint, so the $40$ drift values are not
mutually independent. The count-level bootstrap does not model this
shared-$G_2$ correlation, so the CIs are likely optimistic; we read
the sweep trend, not any single interval, as load-bearing.

\paragraph{Protocol caveats.} Each of the following is load-bearing:
$n_b=200$ fixed sample, greedy decoding, NF4 quantisation, the
count-level reference null, an English-only basket, Kendall-$\tau$ low
power at $|\mathcal{F}|=4$, the median-threshold
$\hat{p}_{\mathrm{flip}}$, and contamination not separately audited.
The claim is non-transferability within a named open-source line under
this protocol; it does not extend, without re-measurement, to
canonical protocols, closed APIs, or larger models.

\section{Conclusion}
\label{sec:conclusion}

We audited four open-source chat-LLM release lines (Yi, Qwen,
Mistral, Gemma) at successive public generations on a fixed
$200$-item basket of five non-canonical/proxy trust-benchmark variants
under three prompt templates. \textbf{Within this protocol}, mean
absolute adjacent-generation Score Drift Rate is several times a
count-level independence reference null and stays in $[6.36,
9.43]$\,pp under every perturbation (\S\ref{find:1},
Table~\ref{tab:robustness}). We report score \emph{non-transferability
across adjacent releases of a named open-source line} as a robust
descriptive effect, not a formal significance claim; the exploratory
diagnostics are not load-bearing. \textbf{What this supports} is
narrow: stamp the checkpoint identifier and evaluation date on the
model card, and re-audit on each materially new release rather than
carry an older score forward. \textbf{Out of scope:} closed APIs,
larger models, canonical protocols, month-cadence rules, and
item-subset uncertainty (multi-seed item resampling is the natural
next check, \S\ref{sec:limits}).

\IEEEtriggeratref{16}
\IEEEtriggercmd{\newpage}
\bibliographystyle{IEEEtran}
\bibliography{references}

@IEEEtranBSTCTL{IEEEexample:BSTcontrol,
  CTLuse_url = "no"
}

@inproceedings{DBLP:conf/acl/LinHE22,
  author       = {Stephanie Lin and
                  Jacob Hilton and
                  Owain Evans},
  editor       = {Smaranda Muresan and
                  Preslav Nakov and
                  Aline Villavicencio},
  title        = {TruthfulQA: Measuring How Models Mimic Human Falsehoods},
  booktitle    = {Proceedings of the 60th Annual Meeting of the Association for Computational
                  Linguistics (Volume 1: Long Papers), {ACL} 2022, Dublin, Ireland,
                  May 22-27, 2022},
  pages        = {3214--3252},
  publisher    = {Association for Computational Linguistics},
  year         = {2022},
  url          = {https://doi.org/10.18653/v1/2022.acl-long.229},
  doi          = {10.18653/V1/2022.ACL-LONG.229},
  bibsource    = {dblp computer science bibliography, https://dblp.org}
}

@inproceedings{DBLP:conf/acl/ParrishCNPPTHB22,
  author       = {Alicia Parrish and
                  Angelica Chen and
                  Nikita Nangia and others},
  editor       = {Smaranda Muresan and
                  Preslav Nakov and
                  Aline Villavicencio},
  title        = {{BBQ:} {A} hand-built bias benchmark for question answering},
  booktitle    = {Findings of the Association for Computational Linguistics: {ACL} 2022,
                  Dublin, Ireland, May 22-27, 2022},
  series       = {Findings of {ACL}},
  pages        = {2086--2105},
  publisher    = {Association for Computational Linguistics},
  year         = {2022},
  url          = {https://doi.org/10.18653/v1/2022.findings-acl.165},
  doi          = {10.18653/V1/2022.FINDINGS-ACL.165},
  bibsource    = {dblp computer science bibliography, https://dblp.org}
}

@inproceedings{DBLP:conf/acl/HartvigsenGPSRK22,
  author       = {Thomas Hartvigsen and
                  Saadia Gabriel and
                  Hamid Palangi and others},
  editor       = {Smaranda Muresan and
                  Preslav Nakov and
                  Aline Villavicencio},
  title        = {ToxiGen: {A} Large-Scale Machine-Generated Dataset for Adversarial
                  and Implicit Hate Speech Detection},
  booktitle    = {Proceedings of the 60th Annual Meeting of the Association for Computational
                  Linguistics (Volume 1: Long Papers), {ACL} 2022, Dublin, Ireland,
                  May 22-27, 2022},
  pages        = {3309--3326},
  publisher    = {Association for Computational Linguistics},
  year         = {2022},
  url          = {https://doi.org/10.18653/v1/2022.acl-long.234},
  doi          = {10.18653/V1/2022.ACL-LONG.234},
  bibsource    = {dblp computer science bibliography, https://dblp.org}
}

@inproceedings{DBLP:conf/emnlp/NangiaVBB20,
  author       = {Nikita Nangia and
                  Clara Vania and
                  Rasika Bhalerao and others},
  editor       = {Bonnie Webber and
                  Trevor Cohn and
                  Yulan He and
                  Yang Liu},
  title        = {CrowS-Pairs: {A} Challenge Dataset for Measuring Social Biases in
                  Masked Language Models},
  booktitle    = {Proceedings of the 2020 Conference on Empirical Methods in Natural
                  Language Processing, {EMNLP} 2020, Online, November 16-20, 2020},
  pages        = {1953--1967},
  publisher    = {Association for Computational Linguistics},
  year         = {2020},
  url          = {https://doi.org/10.18653/v1/2020.emnlp-main.154},
  doi          = {10.18653/V1/2020.EMNLP-MAIN.154},
  bibsource    = {dblp computer science bibliography, https://dblp.org}
}

@inproceedings{DBLP:conf/naacl/RottgerKVA0H24,
  author       = {Paul R{\"{o}}ttger and
                  Hannah Kirk and
                  Bertie Vidgen and others},
  editor       = {Kevin Duh and
                  Helena G{\'{o}}mez{-}Adorno and
                  Steven Bethard},
  title        = {XSTest: {A} Test Suite for Identifying Exaggerated Safety Behaviours
                  in Large Language Models},
  booktitle    = {Proceedings of the 2024 Conference of the North American Chapter of
                  the Association for Computational Linguistics: Human Language Technologies
                  (Volume 1: Long Papers), {NAACL} 2024, Mexico City, Mexico, June 16-21,
                  2024},
  pages        = {5377--5400},
  publisher    = {Association for Computational Linguistics},
  year         = {2024},
  url          = {https://doi.org/10.18653/v1/2024.naacl-long.301},
  doi          = {10.18653/V1/2024.NAACL-LONG.301},
  bibsource    = {dblp computer science bibliography, https://dblp.org}
}

@article{DBLP:journals/corr/abs-2310-06825,
  author       = {Albert Q. Jiang and
                  Alexandre Sablayrolles and
                  Arthur Mensch and others},
  title        = {Mistral 7B},
  journal      = {CoRR},
  volume       = {abs/2310.06825},
  year         = {2023},
  url          = {https://doi.org/10.48550/arXiv.2310.06825},
  doi          = {10.48550/ARXIV.2310.06825},
  eprinttype   = {arXiv},
  eprint       = {2310.06825},
  bibsource    = {dblp computer science bibliography, https://dblp.org}
}

@article{DBLP:journals/corr/abs-2403-04652,
  author       = {Alex Young and
                  Bei Chen and
                  Chao Li and others},
  title        = {Yi: Open Foundation Models by 01.AI},
  journal      = {CoRR},
  volume       = {abs/2403.04652},
  year         = {2024},
  url          = {https://doi.org/10.48550/arXiv.2403.04652},
  doi          = {10.48550/ARXIV.2403.04652},
  eprinttype   = {arXiv},
  eprint       = {2403.04652},
  bibsource    = {dblp computer science bibliography, https://dblp.org}
}

@article{DBLP:journals/corr/abs-2309-16609,
  author       = {Jinze Bai and
                  Shuai Bai and
                  Yunfei Chu and others},
  title        = {Qwen Technical Report},
  journal      = {CoRR},
  volume       = {abs/2309.16609},
  year         = {2023},
  url          = {https://doi.org/10.48550/arXiv.2309.16609},
  doi          = {10.48550/ARXIV.2309.16609},
  eprinttype   = {arXiv},
  eprint       = {2309.16609},
  bibsource    = {dblp computer science bibliography, https://dblp.org}
}

@article{DBLP:journals/corr/abs-2412-15115,
  author       = {An Yang and
                  Baosong Yang and
                  Beichen Zhang and others},
  title        = {Qwen2.5 Technical Report},
  journal      = {CoRR},
  volume       = {abs/2412.15115},
  year         = {2024},
  url          = {https://doi.org/10.48550/arXiv.2412.15115},
  doi          = {10.48550/ARXIV.2412.15115},
  eprinttype   = {arXiv},
  eprint       = {2412.15115},
  bibsource    = {dblp computer science bibliography, https://dblp.org}
}

@article{DBLP:journals/corr/abs-2403-08295,
  author       = {Gemma Team},
  title        = {Gemma: Open Models Based on Gemini Research and Technology},
  journal      = {CoRR},
  volume       = {abs/2403.08295},
  year         = {2024},
  url          = {https://doi.org/10.48550/arXiv.2403.08295},
  doi          = {10.48550/ARXIV.2403.08295},
  eprinttype   = {arXiv},
  eprint       = {2403.08295},
  bibsource    = {dblp computer science bibliography, https://dblp.org}
}

@article{DBLP:journals/corr/abs-2408-00118,
  author       = {Gemma Team},
  title        = {Gemma 2: Improving Open Language Models at a Practical Size},
  journal      = {CoRR},
  volume       = {abs/2408.00118},
  year         = {2024},
  url          = {https://doi.org/10.48550/arXiv.2408.00118},
  doi          = {10.48550/ARXIV.2408.00118},
  eprinttype   = {arXiv},
  eprint       = {2408.00118},
  bibsource    = {dblp computer science bibliography, https://dblp.org}
}

@article{DBLP:journals/corr/abs-2503-19786,
  author       = {Gemma Team},
  title        = {Gemma 3 Technical Report},
  journal      = {CoRR},
  volume       = {abs/2503.19786},
  year         = {2025},
  url          = {https://doi.org/10.48550/arXiv.2503.19786},
  doi          = {10.48550/ARXIV.2503.19786},
  eprinttype   = {arXiv},
  eprint       = {2503.19786},
  bibsource    = {dblp computer science bibliography, https://dblp.org}
}

@article{DBLP:journals/tmlr/SrivastavaRRSAF23,
  author       = {Aarohi Srivastava and
                  Abhinav Rastogi and
                  Abhishek Rao and others},
  title        = {Beyond the Imitation Game: Quantifying and extrapolating the capabilities
                  of language models},
  journal      = {Trans. Mach. Learn. Res.},
  volume       = {2023},
  year         = {2023},
  url          = {https://openreview.net/forum?id=uyTL5Bvosj},
  bibsource    = {dblp computer science bibliography, https://dblp.org}
}

@inproceedings{DBLP:conf/fat/MitchellWZBVHSR19,
  author       = {Margaret Mitchell and
                  Simone Wu and
                  Andrew Zaldivar and others},
  editor       = {danah boyd and
                  Jamie H. Morgenstern},
  title        = {Model Cards for Model Reporting},
  booktitle    = {Proceedings of the Conference on Fairness, Accountability, and Transparency,
                  FAT* 2019, Atlanta, GA, USA, January 29-31, 2019},
  pages        = {220--229},
  publisher    = {{ACM}},
  year         = {2019},
  url          = {https://doi.org/10.1145/3287560.3287596},
  doi          = {10.1145/3287560.3287596},
  bibsource    = {dblp computer science bibliography, https://dblp.org}
}

@inproceedings{DBLP:conf/emnlp/PerezHSCRAGMI22,
  author       = {Ethan Perez and
                  Saffron Huang and
                  H. Francis Song and others},
  editor       = {Yoav Goldberg and
                  Zornitsa Kozareva and
                  Yue Zhang},
  title        = {Red Teaming Language Models with Language Models},
  booktitle    = {Proceedings of the 2022 Conference on Empirical Methods in Natural
                  Language Processing, {EMNLP} 2022, Abu Dhabi, United Arab Emirates,
                  December 7-11, 2022},
  pages        = {3419--3448},
  publisher    = {Association for Computational Linguistics},
  year         = {2022},
  url          = {https://doi.org/10.18653/v1/2022.emnlp-main.225},
  doi          = {10.18653/V1/2022.EMNLP-MAIN.225},
  bibsource    = {dblp computer science bibliography, https://dblp.org}
}

@article{DBLP:journals/corr/abs-2211-09110,
  author       = {Percy Liang and
                  Rishi Bommasani and
                  Tony Lee and others},
  title        = {Holistic Evaluation of Language Models},
  journal      = {CoRR},
  volume       = {abs/2211.09110},
  year         = {2022},
  url          = {https://doi.org/10.48550/arXiv.2211.09110},
  doi          = {10.48550/ARXIV.2211.09110},
  eprinttype   = {arXiv},
  eprint       = {2211.09110},
  bibsource    = {dblp computer science bibliography, https://dblp.org}
}

@article{DBLP:journals/corr/abs-2407-10671,
  author       = {An Yang and
                  Baosong Yang and
                  Binyuan Hui and others},
  title        = {Qwen2 Technical Report},
  journal      = {CoRR},
  volume       = {abs/2407.10671},
  year         = {2024},
  url          = {https://doi.org/10.48550/arXiv.2407.10671},
  doi          = {10.48550/ARXIV.2407.10671},
  eprinttype   = {arXiv},
  eprint       = {2407.10671},
  bibsource    = {dblp computer science bibliography, https://dblp.org}
}

@article{DBLP:journals/corr/abs-2307-09009,
  author       = {Lingjiao Chen and
                  Matei Zaharia and
                  James Zou},
  title        = {How is ChatGPT's behavior changing over time?},
  journal      = {CoRR},
  volume       = {abs/2307.09009},
  year         = {2023},
  url          = {https://doi.org/10.48550/arXiv.2307.09009},
  doi          = {10.48550/ARXIV.2307.09009},
  eprinttype   = {arXiv},
  eprint       = {2307.09009},
  bibsource    = {dblp computer science bibliography, https://dblp.org}
}

@inproceedings{DBLP:conf/iclr/OrenMCLH24,
  author       = {Yonatan Oren and
                  Nicole Meister and
                  Niladri S. Chatterji and others},
  title        = {Proving Test Set Contamination in Black-Box Language Models},
  booktitle    = {The Twelfth International Conference on Learning Representations,
                  {ICLR} 2024, Vienna, Austria, May 7-11, 2024},
  publisher    = {OpenReview.net},
  year         = {2024},
  url          = {https://openreview.net/forum?id=KS8mIvetg2},
  bibsource    = {dblp computer science bibliography, https://dblp.org}
}

@inproceedings{DBLP:conf/nips/ZhengC00WZL0LXZ23,
  author       = {Lianmin Zheng and
                  Wei{-}Lin Chiang and
                  Ying Sheng and others},
  editor       = {Alice Oh and
                  Tristan Naumann and
                  Amir Globerson and
                  Kate Saenko and
                  Moritz Hardt and
                  Sergey Levine},
  title        = {Judging LLM-as-a-Judge with MT-Bench and Chatbot Arena},
  booktitle    = {Advances in Neural Information Processing Systems 36: Annual Conference
                  on Neural Information Processing Systems 2023, NeurIPS 2023, New Orleans,
                  LA, USA, December 10 - 16, 2023},
  year         = {2023},
  url          = {http://papers.nips.cc/paper\_files/paper/2023/hash/91f18a1287b398d378ef22505bf41832-Abstract-Datasets\_and\_Benchmarks.html},
  bibsource    = {dblp computer science bibliography, https://dblp.org}
}

@inproceedings{Alzahrani2024Benchmarks,
  author    = {Norah Alzahrani and Hisham Abdullah Alyahya and Yazeed Alnumay and
               Sultan Alrashed and Shaykhah Alsubaie and Yusef Almushaykeh and
               Faisal Mirza and Nouf Alotaibi and Nora Al-Twairesh and
               Areeb Alowisheq and M. Saiful Bari and Haidar Khan},
  title     = {When Benchmarks are Targets: Revealing the Sensitivity of Large Language Model Leaderboards},
  booktitle = {Proceedings of the 62nd Annual Meeting of the Association for
               Computational Linguistics (Volume 1: Long Papers), {ACL} 2024,
               Bangkok, Thailand},
  pages     = {13787--13805},
  year      = {2024},
  publisher = {Association for Computational Linguistics},
  doi       = {10.18653/v1/2024.acl-long.744},
  url       = {https://aclanthology.org/2024.acl-long.744/}
}

@inproceedings{Kiela2021Dynabench,
  author    = {Douwe Kiela and Max Bartolo and Yixin Nie and Divyansh Kaushik and
               Atticus Geiger and Zhengxuan Wu and Bertie Vidgen and Grusha Prasad and
               Amanpreet Singh and Pratik Ringshia and Zhiyi Ma and Tristan Thrush and
               Sebastian Riedel and Zeerak Waseem and Pontus Stenetorp and Robin Jia and
               Mohit Bansal and Christopher Potts and Adina Williams},
  title     = {Dynabench: Rethinking Benchmarking in {NLP}},
  booktitle = {Proceedings of the 2021 Conference of the North American Chapter of
               the Association for Computational Linguistics: Human Language
               Technologies (NAACL-HLT)},
  pages     = {4110--4124},
  year      = {2021},
  publisher = {Association for Computational Linguistics},
  doi       = {10.18653/v1/2021.naacl-main.324},
  url       = {https://aclanthology.org/2021.naacl-main.324/}
}

@article{Dehghani2021BenchmarkLottery,
  author  = {Mostafa Dehghani and Yi Tay and Alexey A. Gritsenko and Zhe Zhao and
             Neil Houlsby and Fernando Diaz and Donald Metzler and Oriol Vinyals},
  title   = {The Benchmark Lottery},
  journal = {arXiv preprint arXiv:2107.07002},
  year    = {2021},
  doi     = {10.48550/arXiv.2107.07002},
  url     = {https://arxiv.org/abs/2107.07002}
}

@inproceedings{Raji2021EverythingBenchmark,
  author    = {Inioluwa Deborah Raji and Emily M. Bender and Amandalynne Paullada and
               Emily Denton and Alex Hanna},
  title     = {{AI} and the Everything in the Whole Wide World Benchmark},
  booktitle = {Proceedings of the Neural Information Processing Systems Track on
               Datasets and Benchmarks (NeurIPS Datasets and Benchmarks)},
  year      = {2021},
  url       = {https://datasets-benchmarks-proceedings.neurips.cc/paper/2021/hash/084b6fbb10729ed4da8c3d3f5a3ae7c9-Abstract-round2.html}
}

@inproceedings{Wang2023DecodingTrust,
  author    = {Boxin Wang and Weixin Chen and Hengzhi Pei and Chulin Xie and
               Mintong Kang and Chenhui Zhang and Chejian Xu and Zidi Xiong and
               Ritik Dutta and Rylan Schaeffer and Sang T. Truong and Simran Arora and
               Mantas Mazeika and Dan Hendrycks and Zinan Lin and Yu Cheng and
               Sanmi Koyejo and Dawn Song and Bo Li},
  title     = {{DecodingTrust}: A Comprehensive Assessment of Trustworthiness in {GPT} Models},
  booktitle = {Advances in Neural Information Processing Systems 36 (NeurIPS 2023)
               Datasets and Benchmarks Track},
  year      = {2023},
  url       = {https://proceedings.neurips.cc/paper_files/paper/2023/hash/63cb9921eecf51bfad27a99b2c53dd6d-Abstract-Datasets_and_Benchmarks.html}
}

@inproceedings{Huang2024TrustLLM,
  author    = {Yue Huang and Lichao Sun and Haoran Wang and Siyuan Wu and
               Qihui Zhang and Yuan Li and Chujie Gao and Yixin Huang and
               Wenhan Lyu and Yixuan Zhang and Xiner Li and Hanchi Sun and
               Zhengliang Liu and Yixin Liu and Yijue Wang and Zhikun Zhang and
               Bertie Vidgen and Bhavya Kailkhura and Caiming Xiong and others},
  title     = {Position: {TrustLLM}: Trustworthiness in Large Language Models},
  booktitle = {Proceedings of the 41st International Conference on Machine Learning (ICML)},
  series    = {Proceedings of Machine Learning Research},
  volume    = {235},
  pages     = {20166--20270},
  year      = {2024},
  publisher = {PMLR},
  url       = {https://proceedings.mlr.press/v235/huang24x.html}
}

@inproceedings{Raji2020Accountability,
  author    = {Inioluwa Deborah Raji and Andrew Smart and Rebecca N. White and
               Margaret Mitchell and Timnit Gebru and Ben Hutchinson and
               Jamila Smith-Loud and Daniel Theron and Parker Barnes},
  title     = {Closing the {AI} Accountability Gap: Defining an End-to-End Framework
               for Internal Algorithmic Auditing},
  booktitle = {Proceedings of the 2020 Conference on Fairness, Accountability, and
               Transparency (FAT* '20)},
  pages     = {33--44},
  year      = {2020},
  publisher = {ACM},
  doi       = {10.1145/3351095.3372873},
  url       = {https://doi.org/10.1145/3351095.3372873}
}

@inproceedings{Sculley2015TechnicalDebt,
  author    = {D. Sculley and Gary Holt and Daniel Golovin and Eugene Davydov and
               Todd Phillips and Dietmar Ebner and Vinay Chaudhary and
               Michael Young and Jean-Fran{\c{c}}ois Crespo and Dan Dennison},
  title     = {Hidden Technical Debt in Machine Learning Systems},
  booktitle = {Advances in Neural Information Processing Systems 28 (NIPS 2015)},
  pages     = {2503--2511},
  year      = {2015},
  url       = {https://proceedings.neurips.cc/paper/2015/hash/86df7dcfd896fcaf2674f757a2463eba-Abstract.html}
}

@article{Kreuzberger2023MLOps,
  author  = {Dominik Kreuzberger and Niklas K{\"u}hl and Sebastian Hirschl},
  title   = {Machine Learning Operations ({MLOps}): Overview, Definition, and Architecture},
  journal = {IEEE Access},
  year    = {2023},
  volume  = {11},
  pages   = {31866--31879}
}


\end{document}